# Recursive Whitening Transformation for Speaker Recognition on Language Mismatched Condition

*Suwon Shon[1], Seongkyu Mun[2], Hanseok Ko[1]*

[1]School of Electrical Engineering, Korea University, South Korea
[2]Dept. of Visual Information Processing, Korea University, South Korea

swshon@korea.ac.kr, hsko@korea.ac.kr

## Abstract

Recently in speaker recognition, performance degradation due to the channel domain mismatched condition has been actively addressed. However, the mismatches arising from language is yet to be sufficiently addressed. This paper proposes an approach which employs recursive whitening transformation to mitigate the language mismatched condition. The proposed method is based on the multiple whitening transformation, which is intended to remove un-whitened residual components in the dataset associated with i-vector length normalization. The experiments were conducted on the Speaker Recognition Evaluation 2016 trials of which the task is non-English speaker recognition using development dataset consist of both a large scale out-of-domain (English) dataset and an extremely low-quantity in-domain (non-English) dataset. For performance comparison, we develop a state-of-the-art system using deep neural network and bottleneck feature, which is based on a phonetically aware model. From the experimental results, along with other prior studies, effectiveness of the proposed method on language mismatched condition is validated.

**Index Terms**: speaker recognition, language mismatched condition, whitening transform

## 1. Introduction

Spoken language systems are usually developed and trained with out-of-domain dataset regardless of the target domain to which the system is applied. This is because acquiring the in-domain development dataset and its labels can be expensive or often impossible. Such resource imbalance between out-of-domain and in-domain produces significant performance degradation on the system. Particularly on speaker recognition, it was explored by many researchers after Johns Hopkins University (JHU) hosted the Domain Adaptation Challenge 2013 (DAC13) workshop to study about and find solutions to this issue [1] based on the i-vector approach which is state-of-the-art in the field. Many successful methods have been explored to adapt or compensate the domain mismatched system hyper-parameters (universal background model, total variability matrix, within and across covariance matrices) utilizing unlabeled in-domain dataset [2]–[5] or out-of-domain dataset only without any in-domain dataset [6]–[8]. Specifically, these studies explored channel domain mismatched problem according to DAC13 experimental protocol which defines development domain as mostly landline calls from the Switchboard (SWB) dataset and target domain as mostly cellular calls from the Speaker Recognition Evaluation 2010 (SRE10) dataset.

In 2016, the National Institute of Standards and Technology (NIST) held periodic evaluation of speaker recognition systems, e.g. SRE16 and made the situation worse and more challenging. They focused on language mismatched condition with low-quantity in-domain unlabeled dataset. Language mismatched condition is set up by limiting the development dataset to be composed of a large English language dataset such as SWB, SRE and Fisher English with very small set of in-domain (non-English) unlabeled dataset while the evaluation dataset is spoken in Tagalog, Cantonese, Cebuano and Mandarin. Due to the language mismatch between development and target domain, posteriori probability of Universal Background Model (UBM) is not properly expected along the input utterances and eventually degrades performance of speaker recognition systems. Prior study have considered multilingual dataset augmentation [9] for language mismatched condition. However, application is possible only if sufficient in-domain dataset exists.

In this paper, it is proposed recursive whitening transformation, a simple but powerful method to improve performance on language mismatched condition. Whitening transformation is an essential step for the state-of-the-art i-vector based speaker recognition system. However, because of the mismatches between development and target domain, conventional target domain matched whitening transformation always contains un-whitened residual components on development domain i-vectors. Thus, recursive whitening transformation can be applied to remove the un-whitened residual components on development domain i-vector while target domain i-vector is preserved as whitened.

To verify the proposed approach, experiments were conducted on SRE16 language mismatched condition using state-of-the-art i-vector extraction systems based on Gaussian Mixture Model (GMM), Deep Neural Network (DNN) and Bottleneck feature (BNF).

## 2. Speaker Recognition in Language Mismatched Conditions

### 2.1. Speaker recognition system and adaptation

Fig.1 is a high level flow chart of the i-vector based speaker recognition system indicating the parameters required for each process. The first and second block can be estimated using a large source domain dataset. The third block is the pre-processing step of the i-vector by whitening and length normalizing. The fourth block scores expectation of input utterances from the speaker model using the Probabilistic Linear Discriminant Analysis (PLDA) parameters Within-speaker covariance (WC) and Across-speaker Covariance (AC) [10]–[12].

For best performance on domain mismatched condition, the domain of the data used in system development should match the domain of the system that will be applied to. Garcia-Romero [13] found domain for UBM and total variability have limited effects on performance improvement, and the performance depends heavily on the domain for length normalization, whitening transformation and PLDA parameter estimation. Thus, when adapting the system to a new domain, parameters of the third and fourth block must be estimated on the in-domain, e.g. target domain matched, dataset.

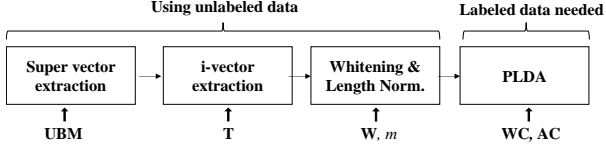

Figure 1 : *Block diagram of conventional speaker recognition system*

### 2.2. SRE16 and language mismatched condition

The dataset for SRE 16 trials are collected from speakers residing outside North America and speaking Tagalog and Cantonese (referred as *major* language) and Cebuano and Mandarin (referred as *minor* language). While both major and minor language have small amount of utterances, especially, minor has extremely small quantity as below in Table 1. For minor and major language set, total 24,140 (4,828 targets and 19,312 non-targets) and 1,986,728 (37,062 targets and 1,949,666 non-targets) trials are composed between enrollment and test utterances, respectively. To set up language mismatched condition, the speaker recognition system was developed using an English language dataset including SWB, Fisher English and previous SRE datasets. Unlabeled datasets of minor and major language are free to use as in-domain dataset for domain adaptation and compensation although they were limited to a small quantity.

Table 1 : *Statistics of SRE16 evaluation dataset.*

| Language set | Category | Labels | Numbers of | | |
|---|---|---|---|---|---|
| | | | Utt. | Spk. | Calls |
| Minor | Enrollment | Available | 120 | 20 | 60 |
| Minor | Test | Available | 1207 | 20 | 140 |
| Minor | Unlabeled | X | 200 | 20 | 200 |
| Major | Enrollment | Available | 1202 | 201 | 602 |
| Major | Test | Available | 9294 | 201 | 1408 |
| Major | Unlabeled | X | 2272 | X | X |

### 2.3. Whitening transformation on domain mismatched condition and un-whitened residual components by language mismatch

Whitening transformation is linear transformation that decorrelates a vector of random variables and forces all variance of dimension to unit variance, so that the covariance of transformed random variable becomes identity matrix. In domain mismatched condition, it is a common approach to get better performance by application of a whitening transformation matrix derived by in-domain dataset although it is unlabeled and small amount of audios are available [2]–[5]. If in-domain dataset is unavailable, sub-corpora label of out-of-domain dataset could be used to compensate the domain mismatches [6]–[8].

Suppose, for SRE16 trials, $x$ be i-vectors from the minor unlabeled, i.e. in-domain, dataset and have a precision matrix and mean of $\mathbf{A}$ and $b$, respectively. Also, let $y$ be i-vectors from the SRE, i.e. out-of-domain, dataset and have a precision matrix and mean of $\mathbf{C}$ and $d$. Let $z$ be the minor enrollment and test i-vectors which has $\mathbf{E}$ and $g$ for precision matrix and mean. Purpose of whitening transformation is to whiten $z$ and $y$ for scoring and PLDA estimation, respectively. As prior studies, in-domain i-vector $x$ can be used for deriving whitening transformation matrix. According to Cholesky whitening transformation, $y$ and $z$ can be whitened using $\mathbf{A}$ and $b$ as below.

$$y' = \mathbf{A}(y - b) \qquad (1)$$

$$z' = \mathbf{A}(z - b). \qquad (2)$$

Then the mean and covariance of $y'$ and $z'$ are

$$\mu(y') = \mathbf{A}(d - b), \qquad Cov(y') = \mathbf{AC}^{-1} \qquad (3)$$

$$\mu(z') = \mathbf{A}(g - b) \approx 0, \qquad Cov(z') = \mathbf{AE}^{-1} \approx \mathbf{I}. \qquad (4)$$

Because $x$ and $z$ is in-domain matched i-vector, it is assumed mean and covariance of $z'$ would be 0 and identity matrix $\mathbf{I}$. Although the out-of-domain $y'$ is still remained as un-whitened because of domain mismatches, prior studies investigated $y'$ is still effective to estimate PLDA parameters and tells $y'$ is close to white rather than the original $y$. For maximum effectiveness, we use out-of-domain sub-corpora dataset to removing un-whitened residual components in out-of-domain i-vector $y'$. Rather than relying on conventional single in-domain whitening transform, we propose recursive whitening transformation approach to remove the un-whitened residual components on language domain mismatched condition by using sub-corpora dataset for whitening transformation sequentially.

## 3. Recursive whitening transformation

A recursive whitening transformation can be performed by the below description with very small in-domain unlabeled dataset and large scale out-of-domain dataset. Let $S_i(j)$ and $\mu_i(j)$ be the $i$-th sub-corpora level, $j$-index sub-corpora precision matrix and mean vector. At each $i$ level, closest sub-corpora $J_i$ can be determined by maximum likelihood of target domain i-vector with the sub-corpora Gaussian models $\theta_{ij}$ as below

$$J_i = \arg\max_{j \in (1,\ldots K)} p(f_{i-1}(\omega) | \theta_{ij}) \qquad (5)$$

where normal distribution $\theta_{ij}=N(\mu_i(j), S_i(j)^{-1})$ and $K$ is total number of sub-corpora at $i$ level. $f_{i-1}(\omega)$ is input i-vector that is recursively whitened at previous level as follows.

$$f_i(\omega) = \eta\big(S_i(J_i) \cdot f_{i-1}(\omega) - \mu_i(J_i)\big) \qquad (6)$$

where $\eta(\cdot)$ is length normalization function [14]. $f_0(\omega)$ is initial i-vector whitened by in-domain dataset as conventional approach at section 2.3.

For example, in visual, we explored SRE 16 minor language set using recursive whitening transformation. Sub-corpora and its level of SRE 16 minor can be represented as below in Table 2.

Table 2 : *sub-corpora level of SRE 16 dataset*

| Sub-corpora Level $i$ | Sub-corpora (sub-corpora index $j$) | Recursively whitened i-vector at each level $i$ |
|---|---|---|
| 0 | Minor unlabeled dataset (*1*) | $f_0(\omega)$ |
| 1 | SRE(*1*), SWB(*2*) | $f_1(\omega)$ |
| 2 | SRE04(*1*), SRE05(*2*), SRE06(*3*), SRE08(*4*), SRE10(*5*), SWB2 p1~p3(*6~8*), SWB2 c1~c2(*9,10*) | $f_2(\omega)$ |

The i-vector was extracted speaker recognition system which was developed using out-of-domain dataset as section 4.1.2. The distribution of original i-vector of minor enrollment and test dataset and contour of equal probability of other sub-corpora distribution are shown at Figure 2. Using minor unlabeled dataset, it is possible to obtain the in-domain whitened i-vector as $f_0(\omega)$ which is identical to the conventional in-domain whitening transformation result.

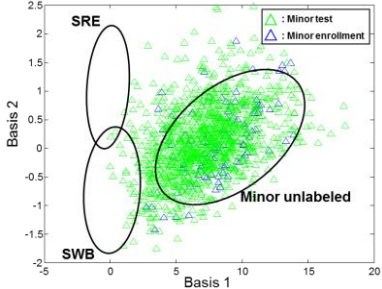

Figure 2 : *Projection of the minor test and enrollment i-vector $\omega$ on 3 dataset PCA subspace (SRE, SWB, minor unlabeled). Ovals represent the equal probability contours of 2-d projection of the SRE, SWB and Minor unlabeled i-vectors. Scatter represents the distribution*

Next, we explored how the dataset is distributed after conventional whitening transformation. Figure 3 represents the whitened minor enrollment and test i-vector $f_0(\omega)$ distribution including contour of SRE and SWB dataset. We could assume $f_0(\omega)$ is white, but out-of-domain i-vector, e.g SRE and SWB, is not as described in Section 2.3. To remove the residual un-whitened components on out-of-domain i-vectors, we could use the SRE dataset ($J_1=1$) for whitening transformation again which is statistically close (by Eq. 5) to minor enrollment and test i-vectors $f_0(\omega)$ to maintain its whitened property.

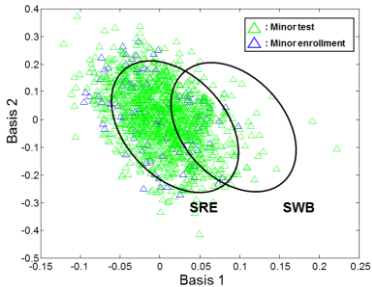

Figure 3 : *Projection of the minor test and enrollment i-vector $f_0(\omega)$ on 2 dataset PCA subspace (SRE, SWB) after whitening transform: The enrollment and test i-vector seems to match with SRE dataset.*

After recursive whitening transformation at sub-corpora level 1, distribution of minor enrollment and test i-vector $f_1(\omega)$ and contour of SRE sub-corpora distribution which consists of SRE04~10 are shown in Figure 4. SRE08 dataset ($J_2=4$) is the statistically closest sub-corpora dataset with enrollment and test i-vector $f_1(\omega)$ distribution. Thus, it can be used as sub-corpora at level 2 recursive whitening transformation to remove the un-whitened residual components in out-of-domain while keeping minor enrollment and test i-vector $f_1(\omega)$ in white. In the rest of the paper, the effectiveness of recursive whitening approach in SRE 16 performance measurement indices is investigated.

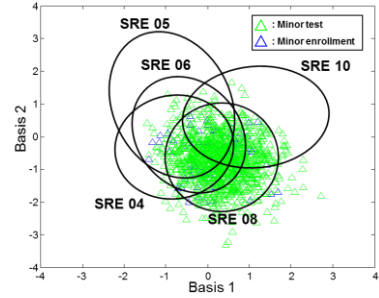

Figure 4 : *Projection of the minor test and enrollment i-vector $f_1(\omega)$ on sub-corpora dataset PCA subspace (SRE 04~10) after whitening transformation twice: The enrollment and test i-vector seems almost whitened already*

## 4. Performance evaluation

### 4.1. Experimental environment

For training speaker recognition system on this paper, Mel-Frequency Cepstral Coefficients (MFCC) is used to generate 60 dimensional acoustic features. It is consisted of 20 cepstral coefficients including log-energy C0, then, it is appended with its delta and acceleration. For training of DNN based acoustic model, different configurations were adopt to generate 40 ceptral coefficient without energy component for high resolution acoustic features (ASR-MFCC). For feature normalization, Cepstral Mean Normalization is applied with 3 seconds-length sliding window. For performance comparison, four different approaches to extract i-vectors are developed as in further sections 4.1.1 to 4.1.4. All i-vector was extracted in 600 dimension. After i-vector extraction and recursive whitening transformation, PLDA parameters were estimated using SRE04~10 dataset for scoring. The number of eigenvoices of PLDA is set to 400.

#### 4.1.1. GMM-UBM

According to general i-vector extraction approach [15], i-vector based speaker recognition system is developed as in Figure 1 based on GMM-UBM. For training of GMM-UBM and total variability matrix, SRE 04~10 and SWB phase2 1~3, cellular 1~2 dataset were used.

#### 4.1.2. DNN-UBM

Fisher English was used for training of Time Delay Neural Network (TDNN) with ASR-MFCC feature. After training TDNN, the DNN-UBM is estimated on high resolution version of MFCC. SRE (04~10, part of 12) and Switchboard Dataset were used for training of DNN-UBM and total variability matrix [16], [17].

#### 4.1.3. Supervised-GMM-UBM (SGMM-UBM)

Phonetically-aware Supervised GMM-UBM [16] was trained using posterior of TDNN network. Same dataset was used for GMM-UBM system to train Supervised GMM-UBM and total variability matrix.

#### 4.1.4. Bottleneck Feature based GMM-UBM (BNF-UBM)

BNF features were extracted using DNN which contains bottleneck layer [18], [19]. DNN layer structure was set to

1500-1500-80-1500 with total 4 layer and MFCC features of all dataset was converted to BNF feature (80 dim) as [20]. After extracting BNF feature, it follows general GMM-UBM based i-vector extraction approaches as in Sec. 4.1.1 and the same dataset was used for GMM-UBM total variability matrix.

### 4.2. Performance comparison on DNN-UBM system

Performance was evaluated in terms of Equal Error Rate (EER), two minimum detection cost $_{min}DCF16-1$ and $_{min}DCF16-2$ with two different cost parameters which are newly determined in SRE16 evaluation plan [21]. The $_{min}C_{primary}$ cost is average of $_{min}DCF16-1$ and $_{min}DCF16-2$.

Performance evaluation was performed on both minor and major languages on SRE 16 dataset as table 3 and 4. After recursive whitening transformation, the system shows considerably higher performance than the single whitening transformation approach. Level 1 recursive whitening provides 16% and 3% improvement in EER, 11% and 6% improvements in $_{min}C_{primary}$ on minor and major trials. For level 2 recursive whitening, it is shown that similar performance is given as level 1 in EER, but it still shows slight improvements in cost indices. In addition, minor language trials show considerable performance improvement compared to major trials. This indicates that the recursive whitening approach is more effective when the in-domain dataset is too small to represent in-domain variability. While recursive whitening transformation shows clear improvements on performance evaluation, the conventional domain compensation techniques such as IDVC [6], DICN [4] that could be applied on this language mismatched condition did not show notable improvements on all indices.

Table 3 : *Performance evaluation on SRE16 minor language (Cebuano and Mandarin) using recursive whitening transformation*

| | Sub-corpora for whitening | | | Compen -sation | EER | $_{min}DCF16-1$ | $_{min}DCF16-2$ | $_{min}C_{primary}$ |
| --- | --- | --- | --- | --- | --- | --- | --- | --- |
| | Level 0 | Level 1 | Level 2 | | | | | |
| Conventional (Level 0 recursive whitening) | Minor | | | - | 21.02 | 0.8217 | 0.8598 | 0.8407 |
| | Minor | | | IDVC | 21.10 | 0.8223 | 0.839 | 0.8306 |
| | Minor | | | DICN | 21.08 | 0.8193 | 0.8722 | 0.8457 |
| Level 1 recursive whitening | Minor | SRE | | - | 17.48 | 0.7358 | 0.7556 | 0.7457 |
| | Minor | SRE | | IDVC | 17.01 | 0.7198 | 0.7504 | 0.7351 |
| | Minor | SRE | | DICN | **17.33** | 0.7204 | 0.7518 | 0.7361 |
| Level 2 recursive whitening | Minor | SRE | SRE-08 | - | 17.92 | **0.7085** | **0.7447** | **0.7266** |
| | Minor | SRE | SRE-08 | IDVC | 17.21 | 0.7123 | 0.7474 | 0.7298 |
| | Minor | SRE | SRE-08 | DICN | 17.33 | 0.7233 | 0.7465 | 0.7349 |

Table 4 : *Performance evaluation on SRE16 major language (Tagalog and Cantonese) using recursive whitening transformation*

| | Sub-corpora for whitening | | | Compen -sation | EER | DCF01 | DCF02 | $_{min}C_{primary}$ |
| --- | --- | --- | --- | --- | --- | --- | --- | --- |
| | Level 0 | Level 1 | Level 2 | | | | | |
| Conventional (Level 0 recursive whitening) | Minor | | | - | 14.10 | 0.7601 | 0.8342 | 0.7971 |
| | Minor | | | IDVC | 14.02 | 0.7547 | 0.8294 | 0.7920 |
| | Minor | | | DICN | 14.13 | 0.7261 | 0.8044 | 0.7652 |
| Level 1 recursive whitening | Minor | SRE | | - | 13.66 | 0.7201 | 0.7766 | 0.7483 |
| | Minor | SRE | | IDVC | 13.74 | 0.7057 | **0.7571** | 0.7314 |
| | Minor | SRE | | DICN | 13.61 | 0.7215 | 0.7764 | 0.7489 |
| Level 2 recursive whitening | Minor | SRE | SRE-08 | - | **13.56** | 0.7224 | 0.7786 | 0.7505 |
| | Minor | SRE | SRE-08 | IDVC | 13.60 | **0.7029** | 0.7572 | **0.7301** |
| | Minor | SRE | SRE-08 | DICN | 13.63 | 0.7203 | 0.7772 | 0.7488 |

### 4.3. Performance comparison on multiple system

For additional analysis, performance was evaluated to verify the proposed approach with multiple state-of-the-art i-vector extraction systems that were reported in prior studies [22]. In this evaluation, symmetric normalization (S-norm) is adopted for score normalization [23]. For S-norm, unlabeled dataset of minor and major were used as imposter utterances for minor and major trials, respectively. For calibration and fusion, simple linear calibration and fusion were conducted by Bosaris Toolkit [24]. Performance was evaluated in $_{min}C_{primary}$ as well as its actual detection cost $_{act}C_{primary}$ with constant threshold as in the SRE16 evaluation plan [21]

From results provided in Table 5, BNF-UBM shows worst performance although we followed best configuration as [20]. On both conventional and proposed approach, it is interesting that UBM based on phonetically aware model such as DNN, supervised GMM and BNF does not have advantages on language mismatched condition although it was reported that high performance is provided with language matched condition. This indicates that the phonetically aware model is not effective on language mismatched condition. All system shows average 13% improvement in all indices after recursive whitening approach is adopted and DNN-UBM shows best performance on both minimum and actual detection costs.

Table 5 : *Performance evaluation on SRE16 minor language (Cebuano and Mandarin) on multiple i-vector extraction system using recursive whitening transformation*

| i-vector Extraction System Name | Conventional (level 0) | | | Level 1 recursive whitening | | |
| --- | --- | --- | --- | --- | --- | --- |
| | EER | $_{min}C_{primary}$ | $_{act}C_{primary}$ | EER | $_{min}C_{primary}$ | $_{act}C_{primary}$ |
| GMM-UBM | 21.91 | **0.8068** | **0.8271** | 18.93 | 0.7155 | 0.7293 |
| DNN-UBM | **21.21** | 0.8267 | 0.8428 | 19.12 | **0.6862** | **0.7043** |
| SGMM –UBM | 21.23 | 0.8099 | 0.8426 | 20.05 | 0.7251 | 0.7461 |
| BNF-UBM | 23.94 | 0.8973 | 0.9215 | 20.19 | 0.7557 | 0.7824 |
| Fusion of 4 sub-systems | 17.01 | 0.7179 | 0.7313 | 15.67 | 0.6478 | 0.6727 |

## 5. Conclusion

Whitening and length normalization is common and essential component of state-of-the-art speaker recognition system. An alternative approach, recursive whitening transformation, is a relatively simple process allowing conventional i-vector extraction systems to deal with the language mismatch between development and target domain dataset. By recursive whitening transformation, the i-vectors of out-of-domain development dataset get whitened gradually to remove un-whitened residual component while i-vector of in-domain target dataset is maintained as whitened. In the experiments on language mismatched condition, the proposed approach indicates its robust performance especially on the challenging condition where in-domain dataset is extremely small. In addition, we validated our approach on several state-of-the-art i-vector extraction systems with the language mismatched condition. It is claimed that recursive whitening transformation is an effective pre-processing step for i-vector and there are possibilities in future studies to conduct compensation on i-vector feature space.

## 6. Acknowledgement

This work was supported by the National Research Foundation of Korea (NRF) grant funded by the Korea government (MSIP) (No. 2017R1A2B4012720). This subject is supported by Korea Ministry of Environment (MOE) as "Public Technology Program based on Environmental Policy".

# 7. References

[1] "JHU 2013 Speaker Recognition Workshop." [Online]. Available: http://www.clsp.jhu.edu/wp-content/uploads/sites/75/2015/10/WS13-Speaker-DAC.pdf.

[2] D. Garcia-Romero, A. McCree, S. Shum, N. Brummer, and C. Vaquero, "Unsupervised Domain Adaptation for I-Vector Speaker Recognition," in *Proceedings of Odyssey - The Speaker and Language Recognition Workshop*, 2014, pp. 260–264.

[3] S. Shum, D. a. Reynolds, D. Garcia-Romero, and A. McCree, "Unsupervised Clustering Approaches for Domain Adaptation in Speaker Recognition Systems," in *Proceedings of Odyssey - The Speaker and Language Recognition Workshop*, 2014, pp. 265–272.

[4] M. H. Rahman, A. Kanagasundaram, D. Dean, and S. Sridharan, "Dataset-invariant covariance normalization for out-domain PLDA speaker verification," in *Interspeech*, 2015, pp. 1017–1021.

[5] A. Kanagasundaram, D. Dean, and S. Sridharan, "Improving out-domain PLDA speaker verification using unsupervised inter-dataset variability compensation approach," in *IEEE ICASSP*, 2015, pp. 4654–4658.

[6] H. Aronowitz, "Inter dataset variability compensation for speaker recognition," in *IEEE ICASSP*, 2014, pp. 4002–4006.

[7] E. Singer and D. A. Reynolds, "Domain Mismatch Compensation for Speaker Recognition Using a Library of Whiteners," *IEEE Signal Process. Lett.*, vol. 22, no. 11, pp. 2000–2003, 2015.

[8] O. Glembek, J. Ma, P. Matejka, B. Zhang, O. Plchot, L. Burget, and S. Matsoukas, "Domain adaptation via within-class covariance correction in i-vector based speaker recognition systems," in *IEEE ICASSP*, 2014, pp. 4060–4064.

[9] A. Misra and J. H. L. Hansen, "Spoken Language Mismatch in Speaker Verification : An Investigation with NIST-SRE and CRSS Bi-Ling Corpora," in *IEEE Workshop on Spoken Language Technology*, 2014, pp. 372–377.

[10] P. Matejka, O. Glembek, F. Castaldo, M. J. Alam, O. Plchot, P. Kenny, L. Burget, and J. Cernocky, "Full-covariance UBM and heavy-tailed PLDA in i-vector speaker verification," in *IEEE ICASSP*, 2011, pp. 4828–4831.

[11] S. Prince and J. Elder, "Probabilistic linear discriminant analysis for inferences about identity," in *International Conference on Computer Vision*, 2007, pp. 1–8.

[12] S. Shon, S. Mun, D. K. Han, and H. Ko, "Maximum likelihood Linear Dimension Reduction of heteroscedastic feature for robust Speaker Recognition," in *IEEE International Conference on Advanced Video and Signal Based Surveillance (AVSS)*, 2015, pp. 1–5.

[13] D. Garcia-Romero and A. McCree, "Supervised domain adaptation for I-vector based speaker recognition," *IEEE ICASSP*, pp. 4047–4051, 2014.

[14] D. Garcia-Romero and C. Y. Espy-Wilson, "Analysis of i-vector Length Normalization in Speaker Recognition Systems.," in *Interspeech*, 2011, pp. 249–252.

[15] N. Dehak, P. J. Kenny, R. Dehak, P. Dumouchel, and P. Ouellet, "Front-End Factor Analysis for Speaker Verification," *IEEE Trans. Audio, Speech, Lang. Process.*, vol. 19, no. 4, pp. 788–798, May 2011.

[16] D. Snyder, D. Garcia-Romero, and D. Povey, "Time delay deep neural network-based universal background models for speaker recognition," in *IEEE Workshop on Automatic Speech Recognition and Understanding (ASRU)*, 2016, pp. 92–97.

[17] F. Richardson, D. Reynolds, and N. Dehak, "A Unified Deep Neural Network for Speaker and Language Recognition," in *Interspeech*, 2015, pp. 1146–1150.

[18] F. Richardson, S. Member, D. Reynolds, and N. Dehak, "Deep Neural Network Approaches to Speaker and Language Recognition," *IEEE Signal Process. Lett.*, vol. 22, no. 10, pp. 1671–1675, 2015.

[19] S. Mun, S. Shon, W. Kim, and H. Ko, "Deep Neural Network Bottleneck Features for Acoustic Event Recognition," in *Interspeech*, 2016, pp. 2954–2957.

[20] A. Lozano-diez, A. Silnova, J. Gonzalez-rodriguez, and C. Republic, "Analysis and Optimization of Bottleneck Features for Speaker Recognition," pp. 352–357, 2016.

[21] "NIST 2016 Speaker Recognition Evaluation Plan." [Online]. Available: https://www.nist.gov/file/325336.

[22] S. Shon and H. Ko, "KU-ISPL Speaker Recognition Systems under Language mismatch condition for NIST 2016 Speaker Recognition Evaluation," *ArXiv e-prints arXiv:1702.00956*, 2017.

[23] S. Shum, N. Dehak, R. Dehak, and J. R. Glass, "Unsupervised Speaker Adaptation based on the Cosine Similarity for Text-Independent Speaker Verification," *Proc. Odyssey - Speak. Lang. Recognit. Work.*, 2010.

[24] N. Brümmer and E. de Villiers, "The BOSARIS Toolkit: Theory, Algorithms and Code for Surviving the New DCF," in *NIST SRE'11 Analysis Workshop*, 2011.